\documentclass[twoside,slac_one]{revtex4}
\usepackage{graphicx}
\usepackage{fancyhdr}
\usepackage{amsmath} 
\usepackage{bm}
\usepackage{amsxtra}
\usepackage{amssymb}
\usepackage{amsthm}
\usepackage{latexsym}
\usepackage{lscape}

\begin{document}

\title{NO$\nu$A Data Acquisition Software System}

%

\author{X. C. Tian, on behalf of the NO$\nu$A Collaboration}
\affiliation{Department of Physics and Astronomy, University of South Carolina, Columbia, SC, USA}

\begin{abstract}
NO$\nu$A is an accelerator-based neutrino oscillation experiment which has a
great potential to measure the last unknown mixing angle $\theta_{13}$,
the neutrino mass hierarchy,  and the CP-violation phase in lepton
sector with 1) 700 kW beam, 2) 14 mrad off the beam axis, 3) 810
km long baseline. The Near Detector on the Surface is fully  functioning
and taking both NuMI and Booster beam data. The far detector
building  achieved beneficial occupancy on April 13. This proceeding will
focus on the DAQ  software system.
\end{abstract}

\maketitle

\thispagestyle{fancy}


\section{Introduction}
\begin{figure}[h]
\centering
\includegraphics[width=1.0\textwidth]{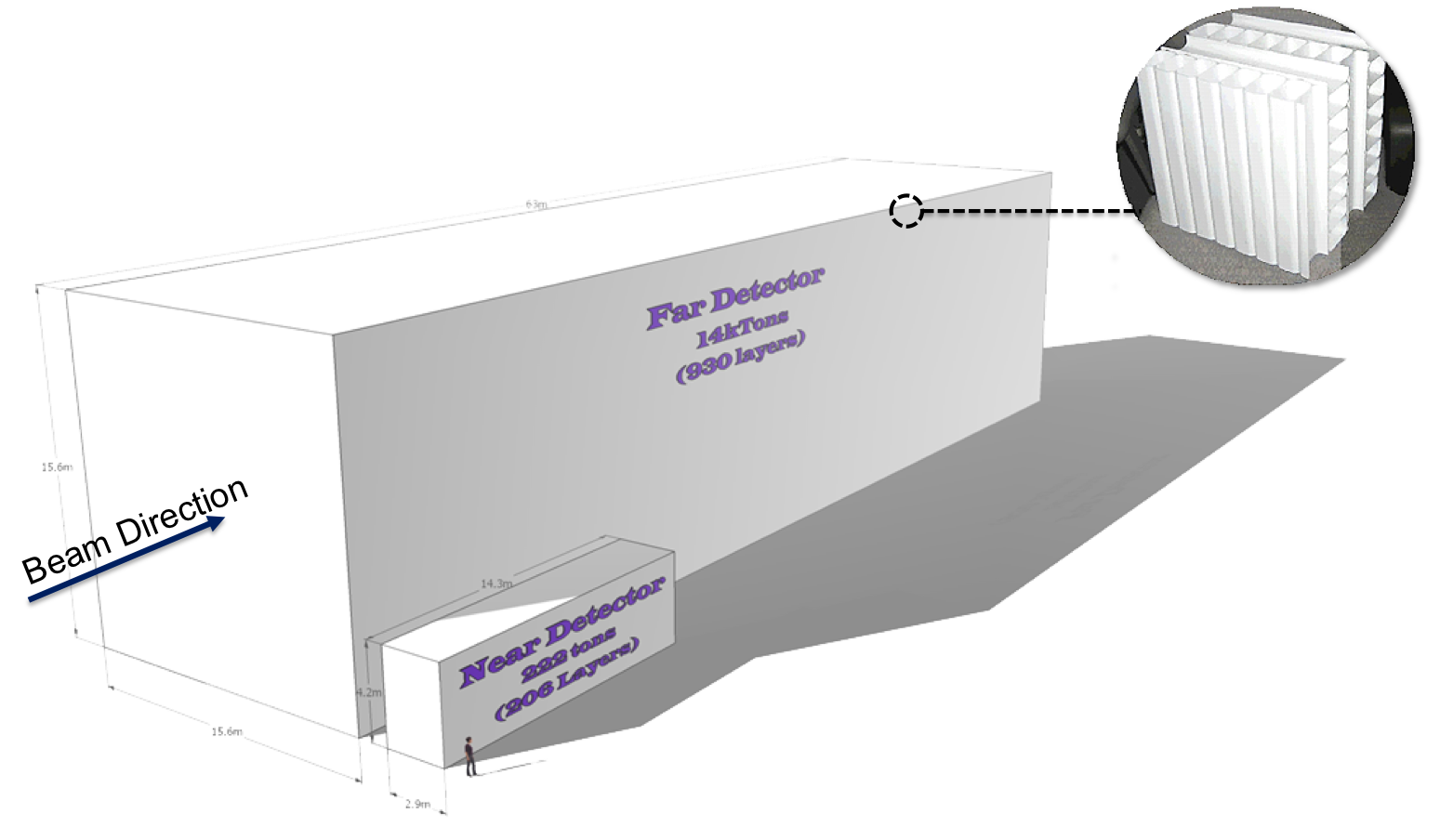}
\caption{The NO$\nu$A detectors. The far (near) detector has 29 (6) blocks and each block is made of 32 scintillator PVC planes, and
928 (192) planes in total.  The near detector also has a muon catcher which is composed of 13 scintillator PVC planes and 10 steel planes.}
\label{detector}
\end{figure}

The next generation long-baseline neutrino experiments~\cite{NOvA, T2K, LBNE} aim to measure the third mixing angle $\theta_{13}$, 
determine whether CP is violated in the lepton sector, and resolve the neutrino mass hierarchy. The NuMI Off-axis electron-neutrino 
($\nu_e$) Appearance (NO$\nu$A) experiment is the flagship experiment of the US domestic particle physics program which has the 
potential to address most of the fundamental questions in neutrino physics raised by the Particle Physics Project Prioritization Panel 
(P5). NO$\nu$A has two functionally identical detectors (Fig.~\ref{detector}), a 222 ton near detector located underground at Fermilab 
and a 14 kiloton far detector located in Ash River, Minnesota with a baseline of 810 km. The detectors are composed of extruded PVC 
cells loaded with titanium dioxide to enhance reflectivity. There are 16,416 and 356,352 cells for the near and far detector, respectively. 
Each cell has a size of 3.93 cm transverse to the beam direction and 6.12 cm along the direction filled with liquid scintillator (mineral 
oil plus 5\% pseudocumene). The corresponding radiation length is 0.15 $X_0$ and the Moliere radius is 10 cm, ideal for the identification 
of electron-type neutrino events. The ``Neutrinos at the Main Injector'' (NuMI) will provide a 14 mrad off-axis neutrino beam to reduce 
neutral current backgrounds and which peaks at 2 GeV, corresponding to the first oscillation maximum for this detector distance. 
The accelerator and NuMI upgrades will double the protons per year delivered to the detector which is $6\times 10^{20}$ protons 
per year. For further details on the current status of NO$\nu$A experiment, please see Ref.~\cite{Gavin}.

\begin{figure}[h]
\centering
\includegraphics[width=\textwidth]{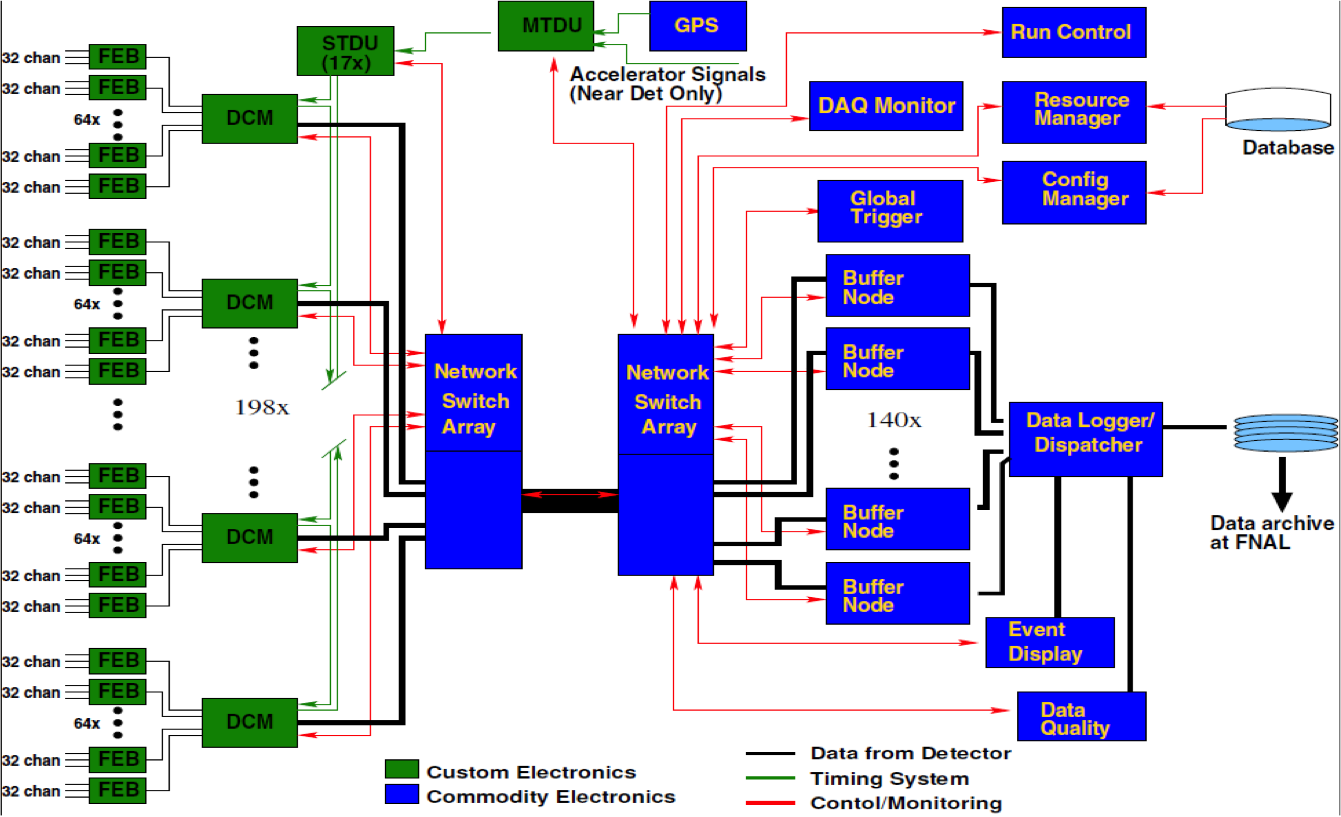}
\caption{An schematic overview of the NO$\nu$A DAQ system. The data stream is from left to right.} \label{daq-all}
\end{figure}

Charged particles from neutrino interactions or cosmic ray muons will emit scintillation light in the scintillator. The scintillation light is 
collected by a loop of wavelength shifting fibers (WLS) and a 32-pixel Avalanche Photo-diode (APD) attached to the fibers converts 
the light pulse into electrical signals. The Data Acquisition (DAQ) System as shown in Fig.~\ref{daq-all} will concentrate the data from 
those APDs into a single stream that can be analyzed and archived. The DAQ can buffer the data and wait for a trigger decision that 
the data should be recorded or rejected. Online trigger processors will be used to analyze the data stream to correlate data with 
similar time stamps and to look for clusters of hits indicating an interesting event.  Additional functionality for dealing with flow control, 
monitoring, system operations and alarms is also included~\cite{NOvA-TDR}.

The event types that NO$\nu$A DAQ will record include beam neutrino events, cosmic ray muons, and other physics events (supernova 
neutrinos, high energy neutrinos, {\it etc.}) Every 2.2 s (will be reduced to 1.3 s with the accelerator and NuMI upgrades), a 10 $\mu$s 
beam spill will be generated and time stamped by a GPS based timing system at Fermilab. All hits that occur in a 30 $\mu$s window 
centered on the 10 $\mu$s spill are recorded for further processing. The event rates are 30 neutrino events per spill for the near detector 
and 1,400 $\nu_e$ beam events per year for the far detector. The randomly selected cosmic ray muons used for calibration and monitoring 
are taken to give 100 times the number of beam neutrino events.  The cosmic ray muon rate is 50 Hz (200 kHz) for the near (far) detector.  Other interesting
physics process such as a supernova explosion at 10 kpc will result in thousands of neutrinos within 10 seconds in the far detector.  For the near
detector, the data rates are 75 TB per year through DAQ system and 1 TB per year written to disk. For the far detector, the data rates are 12,000
TB per year through DAQ sytem and 25 TB per year written to disk.

\section{NO$\nu$A Data Acquisition System}
The primary task for the DAQ is to record the data from APDs for further processing. The data flows through Front End Boards (FEBs), Data 
Concentrate Modules (DCMs), Buffer Nodes (BNs), DataLogger (DL) and then archived on disk or tape as shown in Fig.~\ref{daq-all}. Each 
APD is digitized by a FEB continuously without dead time. The data from a group of FEBs up to 64 are consolidated by the DCM into 5 ms time 
slices which are routed to downstream Buffer Nodes. The data is buffered in the Buffer Nodes for a minimum of 20 s waiting for the spill trigger. 
A spill signal is required to arrive within the buffering time so that the spill time can be correlated with the time-stamped data to determine 
if the hits occurred in or out spill. The triggered data from Buffer Nodes will be merged to form an event in the DataLogger and the event will be 
written to file for storage or shared memory for monitoring. The power distribution system (PDS) provides power to FEBs, APDs, ThermoElectric 
Coolers (TECs)~\footnote{TEC controller cools down the APDs to -15$^\circ$C to keep the noise contribution from photoconversion regions small.}, 
DCMs and Timing Distribution Units (TDUs). The Run Control provides the overall control of the DAQ system. The following sections will describe 
the key subsystems of the NO$\nu$A DAQ system. 

\subsection{Front End Boards (FEBs)}
\begin{figure}[h]
\centering
\includegraphics[width=1.0\textwidth]{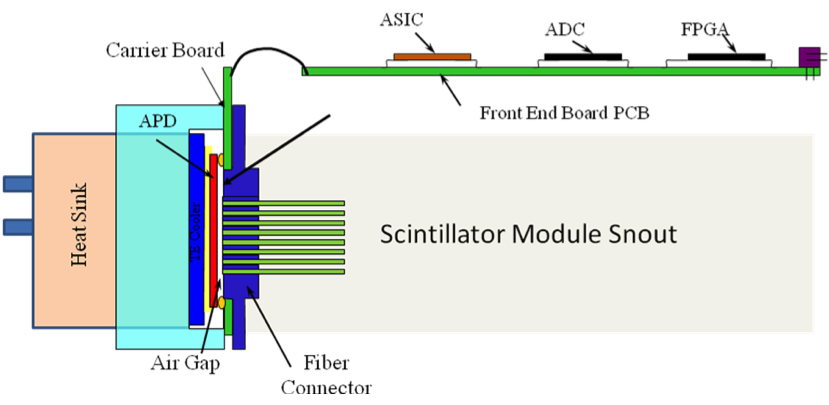}
\caption{Schematic of the APD module and the front-end electronics board showing the major components.  }\label{feb}
\end{figure}

The front end electronics (Fig. ~\ref{feb}) is responsible for amplifying and integrating the signals from the APD arrays, determining the amplitude 
of the signals and their arrival time and presenting that information to the DAQ.  The FEBs are operated in trigger-less, continuous readout mode 
with no dead time, and the data is zero suppressed based on Digital Signal Processing (DSP) algorithms. Data above a pre-programmed threshold 
is settable at the channel level to allow different thresholds to be set depending on the particular characteristics of a given channel. Data above that threshold will be 
time-stamped and compared to a NuMI timing signal in the DAQ system to determine if the event was in or out of spill. Major components of the 
FEB are the carrier board connector location at the left, which brings the APD signals to the NO$\nu$A ASIC, which performs integration, shaping, and 
multiplexing. The chip immediately to the right is the ADC to digitize the signals, and FPGA for control, signal processing, and communication. The ASIC 
is customized to maximize the sensitivity of the detector to small signals from long fibers in the far detector.  The average photoelectrons (PEs) yield at 
the far end of an extrusion module is 30, and the noise is 4 PEs.  The FPGA on the FEB uses a Digital Signal Processing algorithm to extract the time and 
amplitude of signals from the APD. Each FEB reads out 32 channels corresponding to 32 pixels of one APD. Higher detector activity during beam spill at 
the near detector requires higher time resolution, therefore FEBs sample APDs pixels at 8 MHz at the near detector, 2 MHz at the far detector. The FEBs are capable 
of limited waveform digitization and waveform readout.

\subsection{Data Concentrator Modules (DCMs)}
\begin{figure}[h]
\centering
\includegraphics[width=0.9\textwidth]{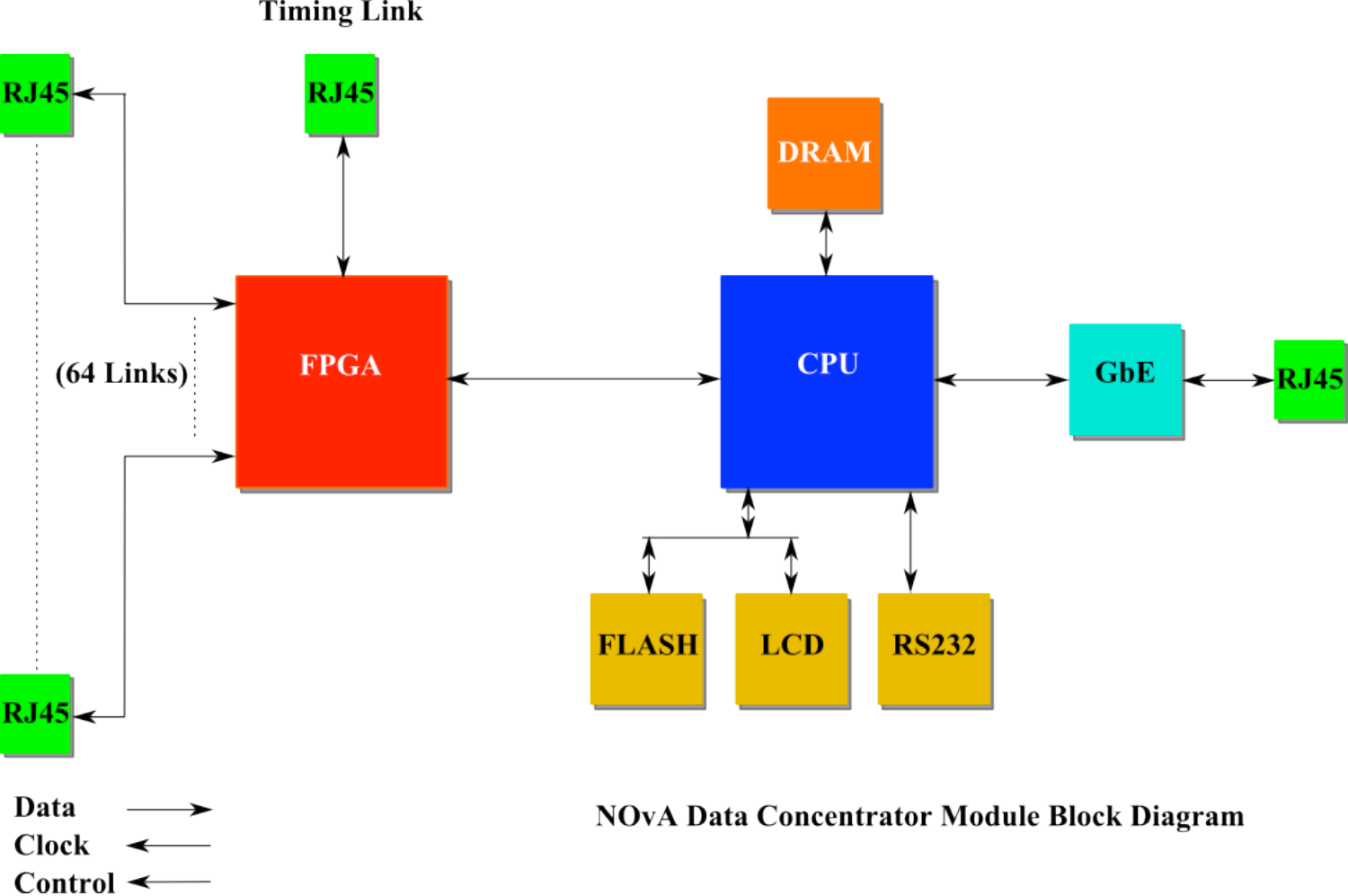}
\caption{The block diagram of the Data Concentrate Module. The DCM FPGA concatenates and 
combines the data received from up to 64 FEBs on dedicated serial links to time slices on the order 
of 50 $\mu$s (MicroSlices). The DCM CPU runs kernel device drivers responsible for the 
programming of the DCM and FEB FPGAs, readout of the MicroSlices and FEB status buffers 
prepared by the DCM FPGA. }\label{dcm}
\end{figure}

The DCM (Fig.~\ref{dcm}) is a custom component of the DAQ. Each DCM is responsible for consolidating the data into 5 ms time slices received from 
up to 64 FEBs to internal data buffers and for transferring the data out to the Event Builder Buffer Farm nodes over Gigabit Ethernet. The 
DCMs also pass timing and control information from the timing system to the FEBs. The DCM mainly consists of a FPGA, an embedded Power PC 
processor, and connectors as shown in Fig.~\ref{dcm}. Data from FEB consists of a header and hit information for a given timeslice. The mid-sized 
FPGA on the DCM concatenates and combines the hit information for all 64 FEBs to time slices on the order of 50 $\mu$s (MicroSlices). The data is 
then read from the FPGA by the embedded Power PC processor running embedded Linux. The DCM application software running on the processor 
has the responsibility of concatenating and packaging the data into 5 ms time slices (MilliSlice) received from the DCM FPGA and distributing the resulting 
data packets to Event Builder Buffer Nodes over a Gigabit Ethernet. The application software also handles communication with Run Control and the DAQ 
Monitor System, provides the high-level interface for programming and configuring the DCM and FEB hardware components, and provides support for 
operating in simulated data input mode. Data is transferred from the DCMs to different Buffer Nodes in a round robin fashion with each DCM starting with 
a different Buffer Node.  Timing information from the timing system is used to divide time into slices such that data packets are transmitted to a different 
Buffer Node in each slice.  In this way, DCMs stay synchronized in their rotation, and no two DCMs ever transmit to the same Buffer Node at once.  Buffer 
Nodes are receiving data from only one DCM at a time, and all Buffer Nodes are receiving data on all time slices. The DCM also connects to the timing 
system which passes a single common timing clock to all FEBs to maintain a sync to common time better than $\pm1$ 64MHz clock tick detector wide.

\subsection{Buffer Nodes (BNs)}
\begin{figure}[h]
\centering
\includegraphics[width=1.0\textwidth,height=4.5cm]{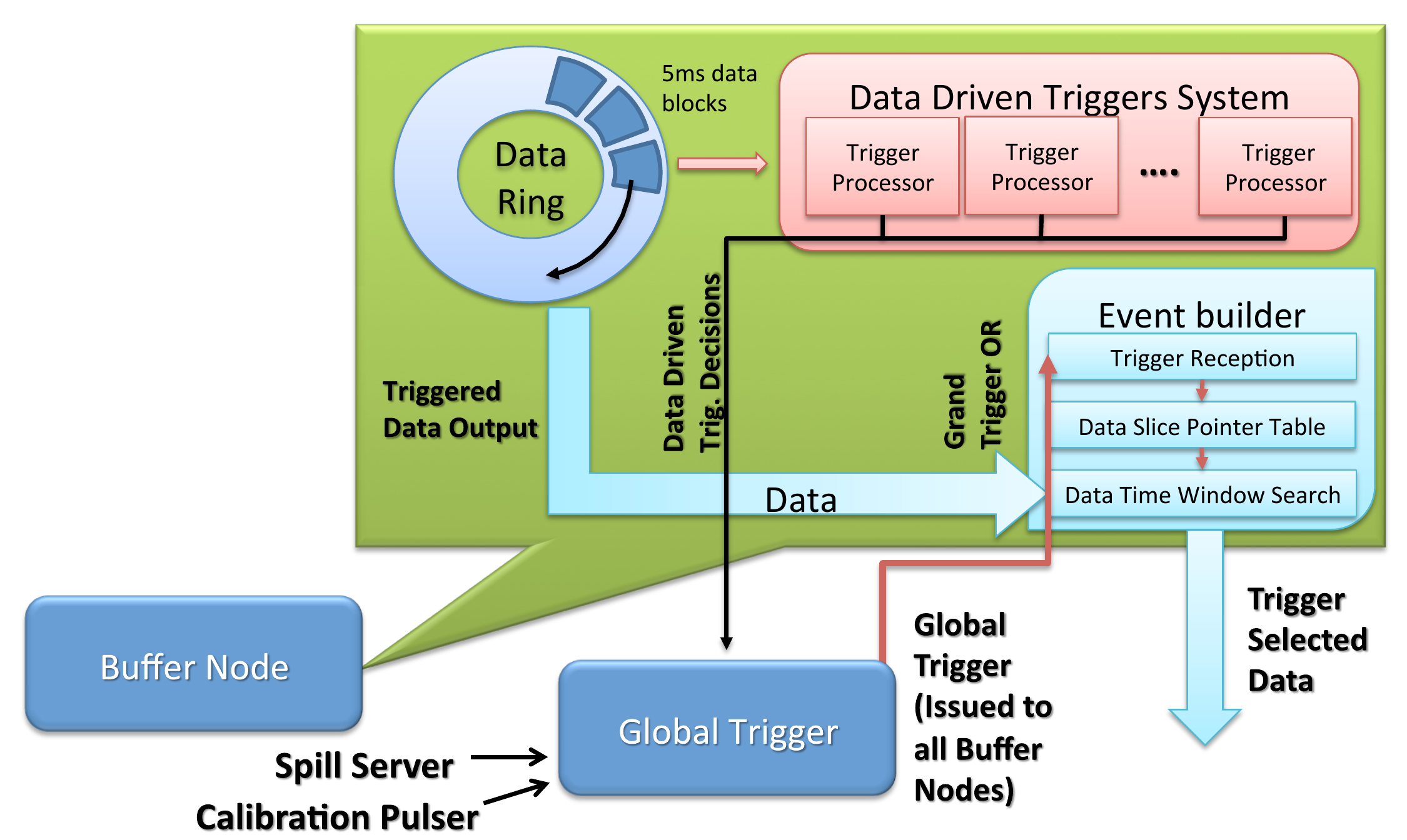}
\caption{The diagram of the Buffer Node. }\label{bn}
\end{figure}

The Buffer Nodes are used to buffer the raw hit data from the entire detector for a minimum of 20 s until a trigger is received from the Global Trigger 
system (Fig.~\ref{bn}). The trigger is a time window and the Buffer Nodes use the time window to search for all data within that frame of time and 
route it to DataLogger process. All data for one time frame will be buffered in the same Buffer Node. Each Buffer Node is independent and contains 
distinct detector data, such that when a given time window is requested, all the Buffer Nodes must perform a search on the available data and at 
least one should return the desired data  The Buffer Node will need to aggregate the data from each of the DCMs for the time frame 
and insure data from all DCMs is seen. This is a form of event building where data for a large time frame is built as a single event. For any frame 
seen by a Buffer Node, data must be received from each DCM. This provides a system integrity check and also allows more performant searching 
when a trigger is received. While all the data from a given time frame will be contained in one farm node, an actual 30 $\mu$s trigger, which defines 
an event, may span these large time frame boundaries and thus be split across two farm nodes. Each farm node will need to determine if it has any 
data for the trigger window in its memory buffer and send it to the logging process. 
\begin{figure}[h]
\centering
\includegraphics[width=1.0\textwidth,height=4.5cm]{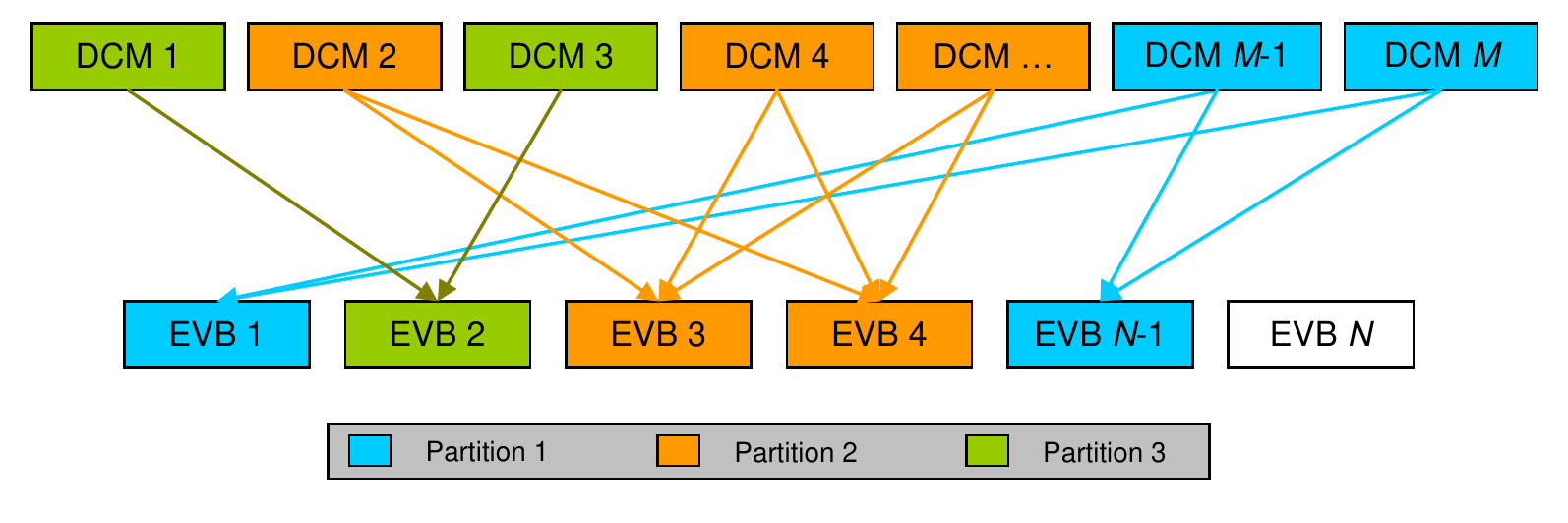}
\caption{The internal structure of the Buffer Node. Data is distributed from the DCMs to the Buffer Nodes in a round-robin fashion with traffic shaping.}\label{dcm-bn}
\end{figure}
	
\subsection{DataLogger (DL)}
\begin{figure}[h]
\centering
\includegraphics[width=1.0\textwidth,height=4.5cm]{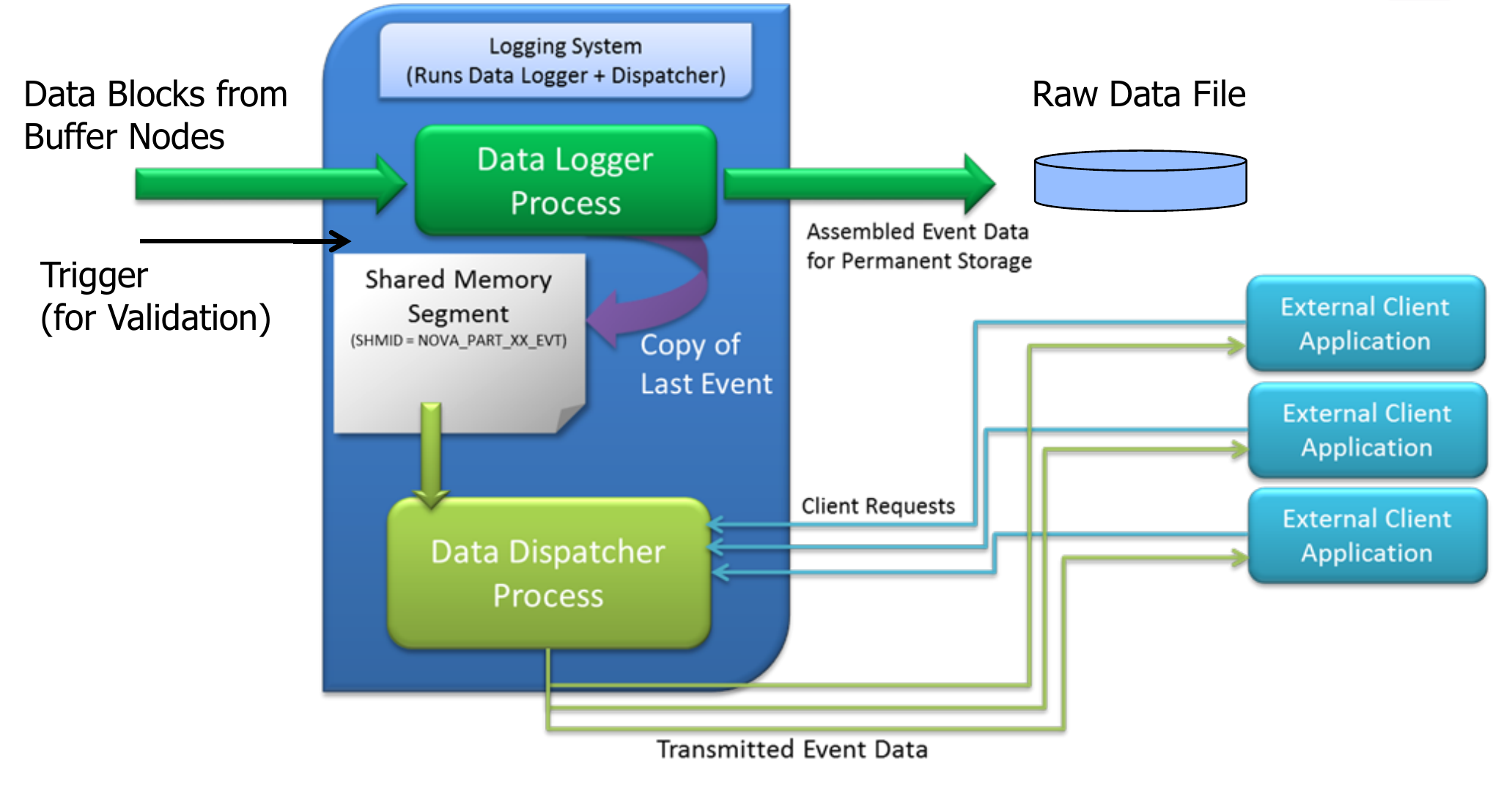}
\caption{The diagram of DataLogger. }\label{dl}
\end{figure}

The DataLogger (DL) is responsible for merging the triggered data blocks from Buffer Nodes to form events and writing them to local 
files or online monitoring systems through a Data Dispatcher process (Fig.~\ref{dl}).  It receives input from the Global Trigger and the Buffer 
Nodes. It writes completed events to run/subrun output disk files and continuously populates a shared memory segment with events for 
online monitoring. The incomplete events called DataAtoms consist of TriggerBlock coming from Global Trigger and/or 
DataBlocks coming from Buffer Nodes. Each DataAtom, whether it is a TriggerBlock or DataBlocks or a combination of both has an
associated Trigger Number which is used to build events. The complete events are written to disk - data streams are formed corresponding 
to each trigger type for the run plus a stream for all events. Failure to build an event which was incomplete is an indication of a DAQ problem 
and generates a warning message sent to the Run Control, with the resulting incomplete event written to disk. These files consist of various 
headers, data blocks, tails, and checksum words. Events are also written to a shared memory segment which is formed at run setup on the 
DataLogger node. This memory segment is overwritten at a constant rate with new events. An Event Viewer is available to view the data 
structure of these events and an online Event Display can also be run on them.

\subsection{Trigger}
The Global Trigger system is responsible for receiving a beam spill signal, or other triggering conditions, such as the periodic calibration ``Pulser'' trigger
and instructs the NO$\nu$A DAQ's data buffer system to save a set of data, rather than the default buffer action which is to forget the oldest data in 
order to make room for current data. Among the variety of triggers, the most important is the beam spill trigger originating in the NuMI 
beam-line, to allow the beam neutrino induced interactions to be recorded. The beam spill signal will be generated at Fermilab in response to the 
signal firing the kicker magnet.  The actual time will be logged and transmitted to the far detector Global Trigger via the Internet. Upon receipt of the 
beam spill time the Global Trigger system will then generate the time windows based on the signal and send that signal to the buffering processes 
on each of the buffer farm nodes.   In the absence of actual spill triggers the system will generate random triggers to allow for tracking of calibration 
and monitoring of the detector. The calibration source for NO$\nu$A is the copious rate of cosmic rays hitting this large detector on the surface. 
While the data rate of these cosmic rays is too great to consider saving all of them, a subset is needed to calibrate, measure the background, and 
monitor the detector. Thus, a Calibration Trigger will be issued at configurable intervals to save blocks of time as determined by operational 
requirements to contain a useful sample of cosmic ray muons. Additionally, Data Driven trigger processing information will be sent from the 
Buffer Nodes to the Global Trigger, which will use that information to decide if the appropriate Data Driven trigger will be issued.  

\subsection{Timing System}
The timing system consists one Master Timing Distribution Unit (MTDU) and serval Slave Timing Distribution Units (STDUs) which is
used to to synchronize both near and far detector timing systems to a known time standard. MTDU is cabled to their 
corresponding STDUs in a daisy-chained fashion, and similarly each STDU will be cabled to the 12 DCMs on its double block in a 
daisy-chain configuration. MTDU derives clock from Global Positioning System (GPS) and distributes to the first STDU in chain.  All 
the FEBs/DCMs are synchronized to this common high precision 16 MHz clock reference which is distributed by STDU. MTDU also 
generates command and sync signals as directed by Run Control, and synchronizes all detector DAQ components on both near and far 
detectors.  The Near Detector MTDU differs from the Far Detector MTDU in that it also connects to the Main Injector timing system, 
and synchronizes itself to the Main Injector. The timing system is self-compensating for cable/transmission propagation delays between 
units (timing units and DCMs). 
\subsection{Message Systems}
The message systems include message passing and message facility. The message passing system which 
is capable of high message bandwidth will be used to transport control and monitoring messages between processes in the DAQ. 
It will support the sending of messages to individual processes as well as groups of processes, and it will provide support for reply 
messages that are generated in response to request messages. General message passing is handled using the FNAL ``Responsive 
Message System (RMS)'' which uses OpenSplice DDS for low level message transmission. The NO$\nu$A specific layers provide 
ease-of-use. 

The message facility system will provide the infrastructure for all of the distributed processes in the DAQ to report status messages of 
various severities in a consistent manner to a central location, and it will provide the tools for displaying and archiving the messages
for later analysis. The message facility system should act as a gate keeper to the Run Control clients with respect to error messages. 
It is the responsibility of this part of the system to determine which messages need to be seen by Run Control. The archiving of status 
messages for later analysis will also be valuable for diagnosing problems after they have occurred.

\subsection{Run Control}
The Run Control system provides a graphical interface for operators to control data taking, control logic to carry out the operators' requests, 
and monitoring functionality to automatically react to exceptional conditions. It is written in C++ using QT~\cite{QT} technology 
and client/server model. All DAQ components implement a well defined state model and under the command of Run Control, make transitions 
between states. The Run Control application must communicate with other applications, including the run history and configuration 
databases, trigger, message systems, and the detector control system. For debugging and commissioning purposes of the large detector 
of essentially one type of detector sub-system, the Run Control will support partitioning of the resources. There will need to be one central resource 
manager that tracks assignments of DCMs and Buffer Nodes to partitions. The resource manager will provide the Run Control client the necessary 
information for using only the hardware and applications reserved for its partition. Run Control will interact with the configuration database to provide 
appropriate information to processes on how to configure themselves, download or receive calibration parameters, {\it etc.}  In addition, it is responsible 
for saving all relevant information on the configuration of the run, including partition information, to a run history database for offline access.  Any changes 
in the configuration would require a new run to be started.  

\subsection{Monitoring Systems}
The DAQ is monitored at different levels by five different applications. The real time monitors include: 1) DAQ Monitor monitors the DAQ 
health and performance using Ganglia~\cite{Ganglia}; 2) Memory Viewer displays the bytes of raw data; 3) Online Event Display displays 
reconstructed events; 4) Online Monitor monitors the run metrics. The offline monitor is data check which checks the detector performance
by looking at metrics of multi-run periods. Please refer to~\cite{Susan} for details.

\section{DAQ Performance on Near Detector On the Surface (NDOS)}
\begin{figure}[h]
\centering
\includegraphics[width=0.48\textwidth]{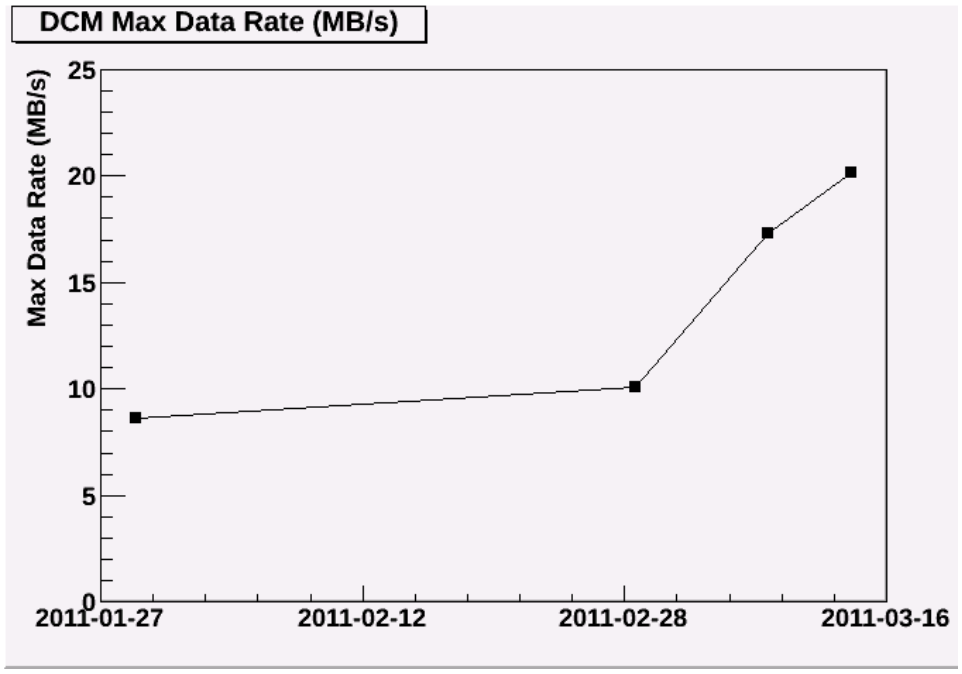}
\includegraphics[width=0.48\textwidth]{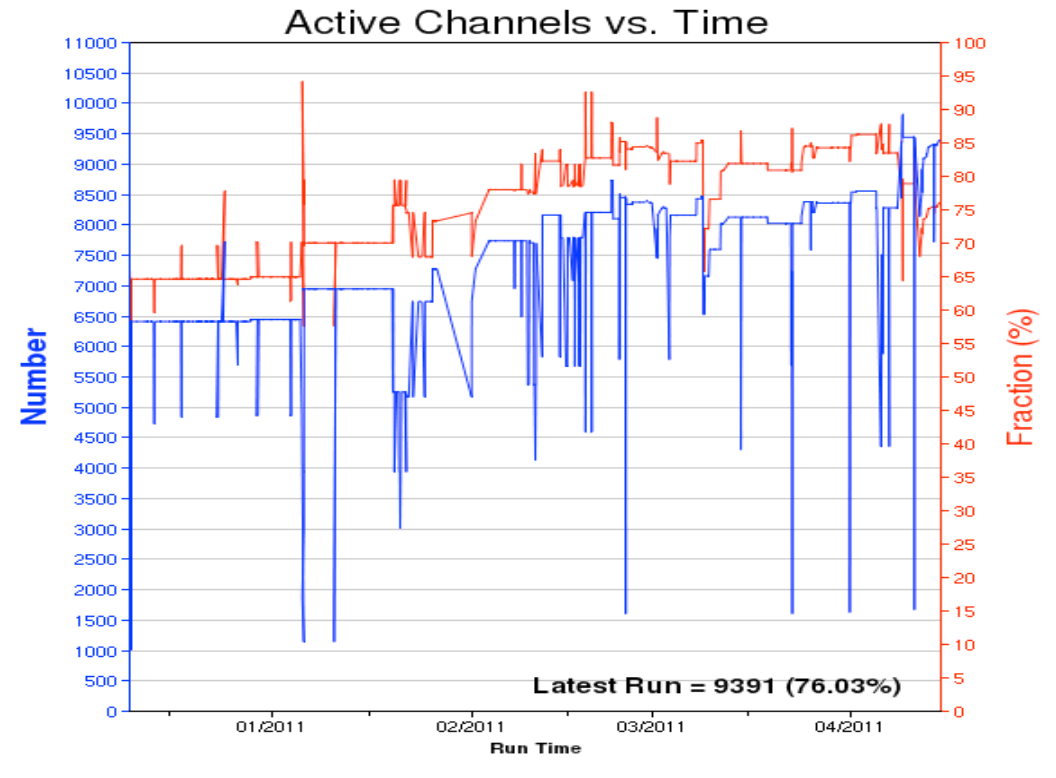}
\caption{The Data Concentrate Module gain as a function of time (left) and the number of active channels as a function of time (right).}\label{dcmgain}
\end{figure}

The Near Detector On the Surface (NDOS) DAQ system has been up and running since fall of 2010. The first neutrino candidate was observed 
on Dec.15, 2010. The NDOS has recorded 8.4$\times10^{18}$ Protons On Target (POTs)  from NuMI neutrino beam, and 5.6$\times10^{19}$ 
POTs from NuMI anti-neutrino beam.  NDOS also accumulated 3.0$\times10^{19}$ POTs  from Booster neutrino beam. Please refer to~\cite{Minerba}
for details. Many DAQ performance gains and fixes of bugs have resulted from NDOS commissioning and running, for example, the DCM data 
through-put has gained a factor of two as a result of optimizing software and network (Fig.~\ref{dcmgain} left). The number of active channels 
(Fig.~\ref{dcmgain} right), live time and quality of data continue to improve over time. 

\section{Conclusion}
Much of the DAQ system has been designed, implemented and deployed to the NO$\nu$A prototype detector, 
Near Detector on the Surface (NDOS) which has been up and running from the Fall of 2010. The system is robust 
enough that can run the readout for the NDOS at near 100\% live readout to disk. Many performance gains and
fixes of bugs have been made as a result of commissioning the Prototype Detector. The far detector will start 
construction starting from Jan. 2012 and expect to readout the first di-block early next year. The far detector DAQ 
system is under development and the system is ready for the rate conditions that will see at the far detector early next year.



\begin{acknowledgments}
The author would like to thank the NO$\nu$A Collaboration for the help on preparing the presentation and proceeding.
\end{acknowledgments}

\bigskip 

\begin{thebibliography}{9}   

\bibitem{NOvA} D. S. Ayres {\it et al.} (NO$\nu$A Collaboration), arXiv:hep-ex/0503053.

\bibitem{T2K}  Y. Itow {\it et al.} (T2K Collaboration), arXiv:hep-ex/0106019.

\bibitem{LBNE}  M. C. Sanchez {\it et al.} (LBNE DUSEL Collaboration), {\it AIP Conf. Proc.} {\bf 1222}, 479 (2010).

\bibitem{Gavin} G. S. Davies (NO$\nu$A Collaboration), DPF proceeding.


\bibitem{NOvA-TDR} ``Data Acquisition System'', Chapter 15, NOvA Technical Design Report.









\bibitem{QT} http://qt.nokia.com/

\bibitem{Ganglia} http://ganglia.sourceforge.net/

\bibitem{Susan} S. Lein (NO$\nu$A Collaboration), DPF proceeding.

\bibitem{Minerba} M. Betancourt (NO$\nu$A Collaboration), DPF proceeding.

\end{thebibliography}

\end{document}